\documentclass[onecolumn]{aa}

\begin{document}

\title{Tunguska genetic anomaly and electrophonic meteors}
 
\author{Z.K. Silagadze}

\institute{Budker Institute of Nuclear Physics, 630 090, Novosibirsk,
              Russia \\
              \email{silagadze@inp.nsk.su}
             }

\date{Received           / accepted           }

\abstract{
One of great mysteries of the Tunguska event is its genetic impact.
Some genetic anomalies were reported in the plants, insects and people
of the Tunguska region. Remarkably, the increased rate of biological
mutations was found not only within the epicenter area, but also along
the trajectory of the Tunguska Space Body (TSB). At that no traces of
radioactivity were found, which could be reliably associated with the 
Tunguska event. The main hypotheses about the nature of the TSB,
a stony asteroid, a comet nucleus  or a carbonaceous chondrite, readily
explain the absence of radioactivity but give no clues how to deal with
the genetic anomaly. A choice between these hypotheses, as far as the
genetic anomaly is concerned, is like to the choice between ``blue devil, 
green devil and  speckled devil'', to quote late Academician 
N.V. Vasilyev. However, if another mysterious phenomenon, electrophonic 
meteors, is evoked, the origin of the Tunguska genetic anomaly becomes 
less obscure.
   }
   
   \maketitle
\section{Introduction}
Tunguska - the scent of mystery and adventure, all over the ninety five 
years. Many theories to what happened many years ago in remote 
region of the Sleeping Land - this is the meaning of the Tartar word Siberia 
( Gallant \cite{gallant}). And none of them can explain all the facts. Not 
very surprising - the systematic research had begun with significant delay
and the facts found are indeed perplexing (Vasilyev \cite{vasilyev},
Bronshten \cite{bronshten}, Zolotov \cite{zolotov}, Zhuravlev \& Zigel
\cite{zhuravlev}, Ol'khovatov \cite{olkhovatov}). What is really surprising
is that this at first glance purely scientific problem raised so much 
interest. It seems the irrational roots of this phenomenon are not always
recognized and appreciated. We are tempted to give some thought to this
side of the Tunguska problem before we embark on the more conventional
scientific track.
 
The truth is that both the public interest in the Tunguska catastrophe and
its scientific exploration were spurred by Kazantsev's (\cite{kazantsev}) 
fantastic suggestion that a nuclear-powered alien spacecraft catastrophe 
caused the Tunguska event (Plekhanov \cite{plekhanov}, Baxter \& Atkins 
\cite{baxter}). Another truth is that the scientific community is rather
reluctant about alien spacecrafts and other UFOs unlike their wordly fellows.
But doing so the scientific community misses one important point: the birth
and rise of the modern age UFO myths as well as their apparent impact on the
popular culture are awesome phenomena begging for scientific explanation. To
our knowledge, Jung (\cite{jung}) was the first to realize scientific 
importance behind seemingly absurd UFO accounts. 

According to Jung these accounts are just a projection of the inner psychic 
state of modern man into the heavens and represent his longing for 
wholeness and unity in this divided, hostile and alien new world. 
Therefore an important message behind of such UFO myths is that they signal 
an increasing psychological stress in the society, changes of the archetypes, 
or psychic dominants, and possibly indicate the end of an era in history 
and the beginning of a new one (Fraim \cite{fraim}). 

In this respect the mythological impact of 
the Tunguska explosion on the native Evenk people, representatives of the 
different culture, is of great interest. Therefore it is not surprising that 
Floyd Favel, one of Canada's most acclaimed playwrights and 
theater directors, decided to develop his next play, The Sleeping Land, on the 
base of great spiritual significance of the Tunguska event for the Evenk 
people (Gordon \& Monkman \cite{favel}). He has really a good story for the 
play. It starts
``with the battle between two Tunguska Evenk clans. Over the years, their 
feud escalated, both clans using their powerful shamans to curse to the other, 
with evil spirits, misfortune and disease. The hostility between them grew 
until one shaman called upon the Agdy to destroy the hated enemy forever. 
These fearsome iron birds fly above the earth in huge clouds, flapping their 
terrible wings to cause thunder, flashing lightening from their fiery eyes. 
On that sunny morning in June, the sky became black as a never ending legion 
of the fearsome birds swooped low over the unfortunate Shanyagir clan. Their 
devastating blasts of fire blew the Shanyagir's tents up into the air over the 
tree tops. The clan's belongings were destroyed, two hundred and fifty of 
their reindeer vanished without a trace, the ancient forest was flattened in 
every direction, and those who still could, fled in panic. To this day, the 
Evenk believe that only the Agdy can live in the area where explosion took 
place. Only a few will risk visiting. And none will live there.'' 
(Gordon \& Monkman \cite{favel})

Although the cultures are different, this Evenk myth has some resemblance 
with the Sodom and Gomorrah Biblical story of miraculous destruction of these 
cities by the raining down of fire from heaven. One can even think that this
ancient myth was also born due to real cosmic event (Clube \& Napier
\cite{clube}). In Koran, the holy book of Islam, one finds a similar story
(Wynn \& Shoemaker \cite{wynn}) ``about an idolatrous king named Aad who 
scoffed at a prophet of God. For his impiety, the city of Ubar and all its 
inhabitants were destroyed by a dark cloud brought on the wings of a great 
wind.'' This last story has an unexpected and adventurous continuation. In
1932 an eccentric British explorer John Philby (Monroe \cite{philby}),
obsessed by the idea to find Ubar, made an arduous trip into the Empty Quarter 
of southern Saudi Arabia, which is one of the most inaccessible and formidable
deserts of our planet (Wynn \& Shoemaker \cite{wabar}). He really found 
something interesting, the place he dubbed Wabar - fortunate misspelling 
because it was not the Lost City of Koran, but the place of the fierce 
meteorite impact (Wynn \& Shoemaker \cite{wabar}, \cite{wynn}). The real
Ubar city was allegedly found much later and this is another breathtaking 
adventure (Clapp \cite{clapp}). Radar images from the Landsat and SPOT remote 
sensing satellites, which uncovered old caravan routes, played the crucial
role in this discovery (El-Baz \cite{elbaz}). Evidence indicates that Ubar
was not destroyed from heaven, instead it fell into sinkhole created by the
underground limestone cavern collapse. But the Wabar meteorite was certainly
capable to destroy Ubar or any other ancient city, because the 12 kilotons 
blast was comparable to the Hiroshima bomb (Wynn \& Shoemaker \cite{wynn}).
The Tunguska explosion was thousand times more powerful, capable to destroy
any modern city. Therefore we come to the conclusion that the unconscious
fears of modern man about hazards from the outer space are not completely
groundless, although not aliens but minor space bodies cause the peril.
It is clear that to reliable estimate this danger it would be helpful to
understand the nature of the Tunguska Space Body (TSB). And this is the point
we embark on the more conventional scientific track, as promised.

There are two main hypotheses on this track about the nature of the TSB:
cometary (Shapley \cite{shapley}, Zotkin \cite{zotkin}, Kresak \cite{kresak},
Fesenkov \cite{fesenkov}) and asteroidal (Kulik \cite{kulik}, Fesenkov 
\cite{fesenkov1}, Sekanina \cite{sekanina}, Chyba et al. \cite{chyba}).
Unfortunately for the science the proponents of these two hypotheses 
practically ignored each other for a long time (Farinella et al. 
\cite{farinella}) assuming the question was settled down once for all by 
their own solution - an interesting example of the Planck's principle 
(Hull et al. \cite{hull}), according to which `` a new scientific truth does 
not triumph because its supporters enlighten its opponents, but because its 
opponents eventually die, and a new generation grows up that is familiar with 
it." There is still no consensus among scientists about the choice between
a comet and an asteroid. Some recent research supports the asteroidal origin 
of the TSB (Foschini \cite{foschini}, Farinella et al.\cite{farinella}) while 
Bronshten (\cite{bronshten1}) advocates the cometary hypothesis indicating 
that despite an extensive and scrupulous search no stony fragments of alleged 
asteroid were found. He gives arguments that neither fireball radiation nor 
air friction can eliminate completely such fragments from the stony asteroid. 

But there are facts which are hard to reconcile with either of these 
hypotheses (Vasilyev \cite{vasilyev2}, Ol'khovatov \cite{olkhovatov}). Below
we discuss genetic impact of the Tunguska event, which is one of such facts.

\section{Biological consequences of the Tunguska event}
Ecological consequences of the Tunguska event have been comprehensively 
discussed by Vasilyev (\cite{vasilyev1}, \cite{vasilyev2}). They constitute
another conundrum of this intricate phenomenon. There were two main types of
effects observed. The first type includes accelerated growth of young and
survived trees on a vast territory, as well as quick revival of the taiga 
after the explosion. The second type of effects is related to the genetic
impact of the Tunguska explosion.

Already participants of Kulik's first expeditions made some observations
about forest recovery in the catastrophe area. In various years the 
impressions were different (Vasilyev \cite{vasilyev1}): in 1929-1930 the  
taiga seemed depressed in this area, while in 1953 no signs of growth
deceleration were seen in comparison with neighboring regions. The first
systematic pilot study of growth of the tree vegetation in the catastrophe
region was performed during 1958 expedition (Vasilyev \cite{vasilyev1}).
Anomalously large tree ring widths up to 9 mm were found in young specimens
which were germinated after the catastrophe, while the average width of the
growth rings before the catastrophe was only 0.2-1.0 mm. Besides the young
trees, the accelerated growth was observed also for the survived old trees.

Stimulated by these first findings, a large scale study of the forest recovery
in the Tunguska area was performed in series of following expeditions after 
1960. In 1968 expedition, for example, morphometric data for more than six 
thousand pine specimens were collected. This vast material establishes the 
reality of the accelerated growth without any doubt (Vasilyev 
\cite{vasilyev1}). More recent study of Longo \& Serra (\cite{longo}) 
confirms this spectacular phenomenon and indicates that the growth has 
weakened only recently for trees of the respectable age of more than 150 years.

The cause of the anomalous growth remains controversial. The most natural and
simple explanation, suggested already in sixties (Vasilyev \cite{vasilyev1}),
assumes that the explosion led to the improved environmental conditions due 
to ash fertilization and decreased competition for light and minerals because 
of the increased distance between trees. Longo \& Serra (\cite{longo}) found
an interesting correlation between the anomalous tree growth and the 
dimensions of the growth rings before the catastrophe. The growth acceleration
was more prominent for trees that grew more slowly before the catastrophe.
But their conclusion that this finding seems to favour the above described
simple hypothesis should be considered as too premature in light of Vasilyev's
(\cite{vasilyev1}, \cite{vasilyev2}) more detailed and broad perspective
analysis of the problem.

According to Vasilyev (\cite{vasilyev1}), an averaging influence 
of the Tunguska event on the final tree dimensions is just a manifestation
of the Wilder's Law of initial values (Wilder \cite{wilder}), which states
that the higher the initial level of some physiological function, the smaller 
the response of the living organism to function-raising agents and the 
greater the response to function-depressing agents, irrespective to the 
stimuli nature. Naturally, the change of the environmental conditions played
a significant role in the taiga recovery. But there are some features of the
accelerated growth phenomenon which are hard to explain solely on the grounds 
of this obvious factor.

The areas where the accelerated growth is observed have different shapes
for the young, after-catastrophe trees and for the old ones that have somehow
survived the catastrophe  (Vasilyev \cite{vasilyev1}). For the young trees
the effect is maximal within the epicenter area. But the region where the
accelerated growth is observed differs significantly both from the area of 
the forest fall and from the area affected by forest fire. This interesting
fact hints that the change in the environmental conditions due to the forest 
devastation is not the leading factor of the accelerated growth in this case. 
Instead, one can suggest that the leading role was played by proximity of
the ancient volcano and the resulting contamination of the soil by volcanic
material (Vasilyev \cite{vasilyev1}). An interesting fact is that the Tunguska
epicenter almost exactly coincides with the muzzle of a Triassic volcano. 
Therefore, if the quick growth of the young trees in the Tunguska area is 
indeed related to the soil enrichment by some rare earth and other elements of 
volcanic origin, it is not surprising that the effect is maximal in the
epicenter area, where the volcano muzzle is also situated. What is surprising 
was found by observing later generation trees. It turned out that the younger
the trees, the higher the concentration of the accelerated growth effect
towards the projection of the TSB trajectory (Vasilyev \& Batishcheva
\cite{batishcheva}, Vasilyev \cite{vasilyev1}). Therefore there should be one 
more factor, directly related to the TSB and possibly of mutagenic nature.

For the old survived trees the effect of the accelerated growth is more
scattered and patchy character. One can find such trees in the forest fall 
area, as well as outside of it. Again, the effect is more prominent in regions
nearby to the TSB trajectory. Besides, the contours of the areas, where the
effect is observed have oval shapes stretched along the direction of the 
TSB trajectory (Emelyanov et al. \cite{emelyanov}, Vasilyev \cite{vasilyev1}).
It is also interesting that there are regions, as for example in the area 
between Kichmu and Moleshko rivers, with considerable forest fall but without
any signs of the accelerated growth among survived trees (Vasilyev 
\cite{vasilyev1}). Moreover, the effect of the accelerated growth does not 
reach its maximum in the investigated area. Instead its extrapolated maximum 
is expected far away from the epicenter, at some 20-25 km distance 
(Emelyanov et al. \cite{emelyanov}, Vasilyev \cite{vasilyev1}). One has an 
impression that the flight of the TSB was accompanied by some unknown agent 
capable to induce remote ecological and maybe even genetic changes.
  
Genetic consequences of the Tunguska event is the most controversial subject.
In sixties some experiments were performed in Novosibirsk to find genetic
effects of ionizing radiation on pines. Among various changes, the most
prominent effect was an increased occurrence of 3-needle cluster pines, while
usually the pine used in experiments had 2-needle clusters. Stimulated by this
finding, G.F. Plekhanov organized special expeditions to study young pines
in the catastrophe area. It turned out that the frequency of 3-needle cluster
trees was really increased in the epicenter area, having the maximum near 
the Mount Chirvinskii - the special point where the TSB trajectory ``pierces''
the Earth's surface and where the effect of accelerated growth also reaches 
its maximum for after-catastrophe trees (Vasilyev \cite{vasilyev1}). However,
as was found later, it is rather common that 3-needle cluster pines occur 
with high frequency in areas with intense forest recovery (after forest fires, 
for example), when pines have large linear increments. Therefore, 
unfortunately, this interesting phenomenon can not be undoubtedly associated 
with the primary factors of the Tunguska explosion and might be a secondary 
effect. 

In seventies V.A. Dragavtsev elaborated a special algorithm to separate  
genotypic and phenotypic variations. Tunguska pine trees linear increments
were processed with this algorithm. It was found that the genotypic dispersion
has sharply increased in the Tunguska trees. The effect is prominent, has a 
patchy character and concentrates toward the epicenter area, as well as 
toward of the TSB trajectory projection (Vasilyev \cite{vasilyev1}, 
\cite{vasilyev2}, \cite{vasilyev}). At maximums the genotypic dispersion shows 
about 12-fold increment (Vasilyev \cite{vasilyev2}). One of the maximums
coincides again with the Mount Chirvinskii, another - with the calculated
center of the light flash (Vasilyev \cite{vasilyev1}).

No indications of increased mutageneses was found in the area, however, in 
later study of pine isozyme systems polymorphysm by electrophoresis method. 
Unfortunately only 11 trees were studied from different locations and the
results were averaged because of sample smallness. Therefore, although this
result does not strengthen Dragavtsev's findings, it is inconclusive to 
reject them either (Vasilyev \cite{vasilyev1}).

Some population-genetic studies were performed in the catastrophe area by 
using a pea Vicia cracca. All studied phenogenetic characteristics were found 
considerably higher in the epicenter area than in the reference point near
the Vanavara settlement (at about 70 km from the epicenter). At that two 
special points with maximal effect were clearly seen in the data. Remarkably,
one of them is again the Mount Chirvinskii and the another one (the Chugrim 
canyon) is only at 1-1.5 km distance from the light flash center (Vasilyev 
\cite{vasilyev1}).

The same researchers studied fluctuating  asymmetry of birch leaves in 
the broader region. It is believed that the fluctuating  asymmetry arises as 
a result of stress the organism experiences during its development and is a 
good measure of its ability to compensate for stress. It was found that the 
asymmetry is significantly increased not only in the epicenter area but also 
in remote regions not affected by the Tunguska explosion (Vasilyev 
\cite{vasilyev1}). This is not surprising because the climatic conditions are 
severe in the Siberian taiga and recent studies indicate that fluctuating  
asymmetry in leaves of birch seems to be a robust indicator of ambient 
climatic stress (Hagen \& Ims \cite{hagen}). Interestingly, in the epicenter 
area one of the sites where the highest asymmetry is observed is noted
Mount Chirvinski (Vasilyev \cite{vasilyev1}). 

In 1969 morphometric peculiarities of ants  Formica fusca were searched in
the epicenter area by inspecting 47 anthills. No noticeable differences were
found at several locations, but ants from the Mount Chirvinskii and from the 
Chugrim canyon were significantly different (Vasilyev \cite{vasilyev1}). 
Unfortunately no reference studies were performed outside the epicenter area.
Analogous studies were continued in 1974-1975 by using ants Formica exsecta.
No peculiarities were found in ants inhabited in the central and peripheral
parts of the catastrophe area (Vasilyev \cite{vasilyev1}).

A very interesting genetic mutation, possibly related to the Tunguska event, 
was discovered by Rychkov (\cite{rychkov}). Rhesus negative persons among the
Mongoloid inhabitants of Siberia are exceptionally rare. During 1959 field
studies, Rychkov discovered an Evenk woman lacking the Rh-D antigen. Genetic
examinations of her family enabled to conclude that a very rare mutation of
the Rh-D gene happened in 1912. This mutation may have affected the women's
parents, who in 1908 lived at some 100 km distance from the epicenter and 
were eyewitnesses of the Tunguska explosion. The women remembered her parents'
impressions of the event: a very bright flash, a clap of thunder, a droning
sound, and a burning wind (Rychkov \cite{rychkov}).

All these facts indicate that the Tunguska event had left a very peculiar
ecological and genetic traces. The hard question, however, is to separate the 
primary and secondary factors leading to the observed phenomena, which may
have complex origins. A recurrent appearance of the TSB trajectory and some
special points related to it in the above given stories suggests nevertheless
that the flight and explosion of the TSB was accompanied by some unknown 
stress factor. A great challenge for the conventional Tunguska theories is 
when to find and explain the nature of this factor. We think that such a 
factor might be electromagnetic radiation. Interestingly, a powerful
electromagnetic radiation is suspected to accompany electrophonic meteors -
an interesting class of enigmatic meteoritic events.

\section{Electrophonic meteors and Tunguska bolide}
The history of the electrophonic meteors research presents another good 
example of the Planck's principle in action. In 1719 eminent astronomer
Edmund Halley collected eyewitnesses accounts on a huge meteor fireball seen 
over much of England. He was perplexed by the fact that many reports declared
the bolide was hissing as if it had been very near from the observer. Being
aware that the sounds can not be transmitted so quickly over distances in 
excess of 100 km, he dismissed the effect as purely psychological, as "the 
effect of pure fantasy". This conclusion and Halley's authority hindered any
progress in the field for two and a half centuries (Keay \cite{keay1}).

At present eyewitnesses accounts on the unusual sounds which accompany some
rare meteoritic events are quite numerous (Vinkovi\'{c} et al. 
\cite{vinkovic}, Keay \cite{keay2}) and almost no more doubts are left about 
reality of the effect. Electrophonic sounds can be divided into two classes
according to their duration. About 10\% of the observed events have short 
duration, about one second, and are of burster type. They produce sharp 
sounds which are reported as ``clicks'' and ``pops''. Other electrophonic
events are of the long duration sustained type and the corresponding sounds
are described as being "rushing" or "crackling" (Keay \cite{keay3}, Kaznev
\cite{kaznev}). Interestingly, similar ``clicks'' have been reported to be 
heard by soldiers during nuclear explosions and it is assumed that they are 
caused by an intense burst of very low frequency (VLF) electromagnetic 
radiation, which is peaked at 12 khz (Johler  \& Monganstern \cite{johler}, 
Keay \cite{keay1}).

The mechanism of how VLF radiation can be generated by a meteoroid was 
proposed by Keay (\cite{keay4}). It is suggested that the geomagnetic field
becomes trapped and ``twisted'' in the turbulent wake of a meteoroid. 
Afterwords the plasma cools and the strain energy of the field is released as 
VLF electromagnetic radiation. The theory was further elaborated by 
Bronshten (\cite{bronshten2}) who showed that as much as megawatt VLF power 
can be easily generated by bright enough bolids.

Extra low frequency (ELF) and VLF electromagnetic fields can be generated
also by other possible mechanisms. For example, explosive disruption of a 
large meteoroid will generate electromagnetic pulses similar to what happens
in nuclear explosions. An electrostatic mechanism of perturbing geoelectric
field, operating for bolids with steep trajectories, was considered by Ivanov 
and Medvedev (\cite{ivanov}). Beech and Foschini developed a space charge
model for electrophonic bursters (Beech \& Foschini \cite{beech1}, 
\cite{beech2}). They suggest that during meteoroid catastrophic breakup a
shock wave is produced which propagates in the plasma around the meteoroid
and leads to a significant space charge due to different  mobility of ions and 
electrons. In this case no significant VLF signal is generated, instead we
have a brief transient in the geoelectric field. 

Suitable transducer is required to transform the VLF energy into audible
form and this is that makes the electrophonic meteor observation such a rare
and capricious phenomenon (Keay \cite{keay1}). From a group of even closely
placed observers one or two may hear the sounds and the others do not. In a 
series of experiments Keay and Ostwald demonstrated  (\cite{keay5}, 
\cite{keay1}) that for audible frequency electric fields various common 
objects can serve as a transducer. For example, volunteers were able to detect
as low as 160 volts peak-to-peak variations of the electric field at 4 kHz 
frequency, with their hair or eyeglass frames acting as a transducer.

Therefore at last we have a clever and scientifically sound explanation of
these mysterious sounds which baffled scholars for centuries. But any theory
needs experimental confirmation. Unfortunately instrumentally recorded 
electrophonic meteor data are very scarce due to extreme rareness of the
phenomenon: by an optimistic prediction a person which would spend every night
outdoors may expect  to hear an electrophonic sound once in a lifetime
(Keay  \& Ceplecha \cite{keay6}, Keay \cite{keay1}). 

In 1993 Beech, Brown and Jones (\cite{beech3}) detected 1-10 kHz broad band 
VLF transient concomitant to the fireball from the  Perseid meteor. However no 
electrophonic sounds were reported. Still earlier a meteor VLF signal was 
detected by Japanese observers (Keay \cite{keay7}). Garaj et al. 
(\cite{garaj}), as well as  Price and Blum (\cite{price}) reported detection
of the ELF/VLF radiation associated with the Leonid meteor storm. A very 
interesting observation was made during reentry of the Russian communications 
satellite Molniya 1-67 (Verveer et al. \cite{verveer}). An observer reported 
an electrophonic sound near the end path of the satellite, which produced a
large orange fireball during its entry in the atmosphere. At the same time
several geophysical stations in Australia detected a distinct ELF (about 1 Hz)
magnetic pulse. Unfortunately no instrumentation was available to detect 
electromagnetic radiation above 10 Hz to check Keay-Bronshten's theory. 
Interestingly, ELF electromagnetic transients may effect the human brain 
directly and therefore may require lower energy levels to produce 
electrophonic effects (Verveer et al. \cite{verveer}). These sparse 
experimental data are clearly insufficient to draw definite conclusions about 
the physics of the radio emissions from meteors (Andrei\'{c}  \& Vinkovi\'{c}
\cite{andreic}). On the other hand, the existing theoretical models are also
too simplified to be able to give a detailed description of the phenomenon
(Bronshten \cite{bronshten3}).

A remarkable benchmark in the research of electrophonic meteors was the first 
instrumental detection of electrophonic sounds during the 1998 Leonid meteor 
shower (Zgrabli\'{c} et al. \cite{zgrablic}). Ironically Leonid meteors are
least suitable devices for production of the VLF radiation via the
Keay-Bronshten mechanism which demands the Reynolds number in the meteor 
plasma flow to exceed $10^6$. In the case of Leonids, which are mostly
dust grains, this leads to unreasonably large initial size $D_0>3~\mathrm{m}$
and mass $\sim 3000~\mathrm{kg}$ (Zgrabli\'{c} et al. \cite{zgrablic}).
Nevertheless two clear electrophonic signals were instrumentally recorded
during the expedition. The first originated from the meteor at the altitude 
of $\sim 110~\mathrm{km}$. The second - at altitude of $85-115~\mathrm{km}$.
In both cases the sounds preceded the meteors' light maximum.  
These features are hard to explain also in other models suggested for 
electrophonic bursters. No ELF/VLF signal was detected in these two events.
But the receiver apparatus was insensitive for frequencies below 500 Hz,
while the frequency range of the observed electrophonic sounds was 37-44 Hz.
If one assumes that these sounds originated from  the transduction of the
ELF/VLF transient, the observed sound intensities will imply unreasonably
high ELF/VLF radiation power, impossible to explain by any theoretical 
mechanism starting from meteor alone (Zgrabli\'{c} et al. \cite{zgrablic}).
Therefore this remarkable observations show that the existing theories are 
at least incomplete and the electrophonic meteor mystery remains still largely
unsolved.  

Zgrabli\'{c} et al. (\cite{zgrablic}) suggested that the Leonids acquire
large enough space charge and can trigger an yet unidentified geophysical 
phenomenon upon entering the E-layer of ionosphere at $\sim 110~\mathrm{km}$.
It is assumed that such phenomena in its turn will generate a powerful EM 
radiation burst. Note that this possibility was advocated by Ol'khovatov 
(\cite{olkhovatov1}) much earlier. 

But Keay-Bronshten mechanism is expected to operate well for slow and bright
bolids (brighter than the full Moon), which penetrate deep into the 
atmosphere. The Tunguska meteorite was clearly of this type. Therefore
we can not exclude that its flight was accompanied by a powerful ELF/VLF
radiation. Are there any eyewitness accounts which support the electrophonic
nature of the Tunguska bolide? 

Already in 1949 Krinov (\cite{krinov}) draw his attention to the strange 
circumstance that many independent observers described sounds which preceded
the appearance of the bolide. He notes that similar phenomena were reported
many times by spectators of electrophonic meteors. But there was a significant
difference: in the Tunguska bolide case the sounds were extraordinary strong, 
more like to powerful strikes than to the feeble electrophonic cracks and 
rustles. Krinov notes further that this difference might be a consequence
of the exceptional magnitude of the Tunguska event. However more likely to
him appears the explanation that all these reports were purely psychological
in nature, that the observers have a unconscious tendency to change the
succession of light and sound effects, or unify them in time by neglecting
the time lag due to low velocity of the sound. Here is one such witness account
(Krinov \cite{krinov}) from K.A. Kokorin, resident of Kezhma village:

``... at hours 8-9 in the morning, not later, the sky was completely clear,
without any clouds. I entered the bath (in the yard) and just succeeded in
taking my shirt out when suddenly heard sounds, resembling a cannonade. At
once I run out in the yard, which had an opened perspective towards the 
south-west and west. The sounds still continued at that time, and I saw in the
south-west direction, at an altitude about the half between the zenith and the
horizon, a flying red sphere with rainbow stripes at its sides and behind it.
The sphere remained flying for some 3-4 seconds and then disappeared in the
north-east direction. The sounds were heard all the time the sphere flew,
but they ceased at once as the sphere disappeared behind the forest.``

Krinov's reaction to this report is very characteristic for the history of 
the electrophonic sounds research. He considers it clearly impossible the 
sounds to precede the bolide flight and concludes that Kokorin just forget
the right succession of the events because of remoteness of the event, 
as the inquiry took place in 1930. Of course in this particular case Krinov
might be right, but the fact that similar assertions can be found in many
other witness accounts forces us to believe just the opposite.

E.E. Sarychev, inquired in 1921 (Vasilyev et al. \cite{vasilyev3}) remembers:
``I was a leather master. In summer (near the spring) at about 8 AM tanners
and I washed wool on the bank of Kana river, when suddenly a noise emerged,
as from wings of a frightened bird, in the direction from south to east,
towards Anzyr village, and a wave went upward the river like ripples. After
this a piercing strike followed, and then dull other strikes, as if from
the underground thunder. The strike was so powerful that one of workers,
Egor Stepanovich Vlasov (who died now), fell into the water. With the
emergence of noise, a radiance appeared in the air, of spherical shape,  
the size of half Moon and with a bluish tinge, quickly flying in the 
direction from Filimonov to Irkutsk. Behind the radiance a trail was being 
left in the form of sky-bluish stripe, stretching along almost all the track 
and gradually decaying from the end point. The radiance disappeared behind 
the mountain without any explosion. I can't notice how long it lasted, but
it appeared only very briefly. The weather was absolutely clear and there was 
the still around''

Ya.S. Kudrin, who was a nine years old child at that time, gives the 
following description of the sounds heard (Vasilyev et al. \cite{vasilyev3}):
``The sound was like a thunder, it ceased after the bolide flight. The sound 
was not very strong, just like ordinary thunder. The sound was moving together
with the object toward the north. The sound was heard before the object 
became visible and it stopped as the object disappeared.''

I.K. Stupin was also 8-10 years old boy in 1908 and also remembers that the 
appearance of the sound preceded the appearance of the object. According to
him the sound was cavernous and of low tone. At that he did not notice any
air wave or vibration of the Earth (Vasilyev et al. \cite{vasilyev3}).

The eyewitness report from V.I. Yarygin is also interesting in this respect
(Konenkin \cite{konenkin}): ''In 1908 I lived in village Oloncovo, at some
35 km from the town Kirensk upstream to Lena river. At that day we rode horses
in a field. At first we heard a loud roar, so that the horses had stopped.
We saw a blackness on the sky, behind of this blackness there were blazing 
tails and then a fog of the black color. The sun vanished and darkness came
down. From this blackness a flame of fire darted from south to north.''

One can easily find witness reports where the character of sounds resembles 
very close the electrophonic ones. For example E.K. Gimmer describes the 
sounds from the meteorite as of sizzling type, as if a red-hot iron was put 
into water (Vasilyev et al. \cite{vasilyev3}). S.D. Permyakov remembers that
there was no roar when the bolide fled above him, instead he heard some 
noise and boom (Konenkin \cite{konenkin}).

Note that the electrophonic nature of the Tunguska bolide was argued earlier
by Khazanovitch (\cite{khazanovitch}), who gives some other examples of 
eyewitness accounts to support this suggestion. Even more earlier Vasilyev
(\cite{vasilyev4}) had discussed an unsolved conundrum that thunderlike 
sounds were heard not only during and after the bolide flight, but also 
before it. He dismissed the psychological explanation of this strange
fact by subjective errors, as the peculiarity had been reported by too many 
independent observers, some of them being at several tens of kilometers from 
the bolide-trajectory projection, and therefore the sounds heard by them can
not be caused by the ballistic wave. He came to the conclusion that the only 
real explanation could be achieved by suggesting a link with powerful 
electromagnetic phenomena, induced by the bolide. 

It is interesting that a terrific roar of presumably electrophonic nature was 
reported by eyewitnesses to accompany overhead passing of a Tunguska-like
bolide in the British Guyana in 1935 (Steel \cite{steel}). Other known 
evidence of electromagnetic disturbances from bolids indirectly supports 
a possibility that the similar phenomena is not excluded also in the 
Tunguska event case. The most recent one is related to the Vitim meteorite 
which fell in Siberia 25th September 2002. The witness report from 
G. K. Kaurtsev, a Mama airport employee, clearly indicates that a strong 
alternate electromagnetic field was induced by the bolide during its flight
which led to the induction phenomenon and to the appearance of St. Elmo's 
fires (Yazev et al. \cite{yazev}): 

`` At night there was no electricity, the 
settlement was disconnected. I woken up and saw a flash in the street. 
The switched off filament lamps of the chandelier lighted dimly, to half their
normal intensity. After 15-20 seconds an underground boom came. Next morning
I went to the dispatcher office of the airport. Security guards Semenova Vera
Ivanovna and Berezan Lydia Nikolaevna have told the following story. They 
were on the beat and have seen that "bulbs were burning" on the wooden poles 
of the fence surrounding the airport's meteorological station. They were 
scared very much. Fires glowed during 1-2 seconds on the perimeter of 
the protection fence. The height of the wooden poles is  approximately 1.5 
meters.'' 

Note that the Mama settlement was at several tens of kilometers distance from 
the bolide's flight path. Besides, the scale of the Vitim event is incomparable
with the scale of the Tunguska explosion: the latter was by about three orders
of magnitude more energetic. Therefore it seems very plausible that the
Tunguska bolide flight might be accompanied by very strong alternate 
electromagnetic fields. The next question is whether these fields could 
lead to the observed ecological and genetic consequences.
 
\section{Biological effects of ELF/VLF electromagnetic radiation}
It is not excluded that one should apply to a magical herb mugwort (Artemisia 
Vulgaris), the  foremost sacred plant among Anglo-Saxon tribes, to resolve 
the enigma of the Tunguska genetic impact. It is said that mugwort can invoke 
prophetic dreams if used in a dream pillow (Cunningham \cite{cunningham}).
So a solution of the riddle could be dreamed up in this way. But jokes aside,
what is really magical about mugwort, it is the suspected ability of Artemisia
Vulgaris to use Earth's magnetic field for adaptation purposes. Mugwort is not
the only plant with such ability. For example, compassplant (Silphium 
Laciniatum) uses the same adaptation strategy, which is even more pronounced 
in this case. Its deeply cut basal leaves are placed vertically and they align 
themselves in a north-south direction. This allows them to avoid excessive 
heat of the overhead midday sun and so minimize the moisture loss, but 
nevertheless have a maximum exposure to the morning and evening sun.

Some animals, including fish, amphibians, reptiles, birds and mammals, 
are also using geomagnetic field for orientation (Wiltschko \& Wiltschko 
\cite{wiltschko}). The sensory basis of magnetoreception is not
completely clear yet. Two types of magnetoreception mechanisms are suspected 
in vertebrates. The evidence for a light-dependent, photoreceptor-based 
mechanism is reviewed by Deutschlander et al. (\cite{deutschlander}) along
with the proposed biophysical models. It is supposed, for example, that 
magnetic field can alter the population of excited states of photosensitive 
molecules, like rhodopsin, which might lead to chemical effects. But some
experiments have shown that light is not necessary for magnetoreception
(Wiltschko \& Wiltschko \cite{wiltschko}). Therefore a mechanism for a direct
sensing of the magnetic field should also exist. This mechanism possibly is
based on chains of single-domain crystals of magnetite in a receptor cell
(Walker et al. \cite{walker}). As the chain rotates in the magnetic field, 
it will open some mechanically gated ion channels in the cell membrane by 
pulling on the microtubule-like strands which connect the channels to the 
chain.

Above given examples indicate that Earth-strength magnetic fields can  
affect biological systems. Moreover, this interaction provides evolutionary
important tools for adaptation. Therefore one can expect that the magnetic 
sense in the biological systems is as perfect as any other known sensory
systems and has evolved down to the thermal noise limit in sensitivity
(Kirschvink et al. \cite{kirschvink}).

Thus it is not surprising that various biological effects of the low
frequency nonionizing electromagnetic radiation have been found, although
the underlying mechanisms responsible for these effects are still not
completely understood (Marino \& Becker \cite{marino}, Binhi \& Savin
\cite{savin}, Becker \& Marino \cite{becker}, Binhi \cite{binhi}). The
potentially hazardous effects of the ELF electromagnetic fields were 
especially scrutinized during last decades, because the power frequencies of 
most nations are in the ELF range. Let us mention some most impressive 
facts from Marino and Becker's review (\cite{marino}).

Even relatively brief exposures to high intensity ELF electric fields were
shown to be fatal to mice, Drosophila and bees. For example, above  
500 v/cm, bees sting each other to death. And 30-500 v/cm at 50 Hz is 
sufficient to change metabolic rate and motor activity.

ELF electric field exposure affects central nervous system. For example,
a significant increase in hypothalamic activity was recorded from the
microelectrodes implanted in anesthetized rats during the 1 h exposure period
to the inhomogeneous electric field of 0.4 v/cm maximum at 640 Hz. Some
in vitro studies indicate effects on the calcium release and biochemical 
function. For example, 1.55 v/cm electric field at 60 Hz caused complete loss 
of biochemical function in brain mitochondria after 40 min exposure.

Exposure to the ELF electric or magnetic field produces a physiological 
stress response. For example, rats exhibited depressed body weights, 
decreased levels of brain choline acetyltransferase activity, and elevated 
levels of liver tryptophan pyrrolase after 30-40 days exposure to 
0.005-1.0 v/cm electric field at 45 Hz.

It was found that an asymmetrically pulsed magnetic field repeating at 
65 Hz with a peak value of several G accelerates the healing of a bone 
fracture in dogs. Some studies indicated a slight enhancement of growth  in 
plants near high-voltage transmission lines. The growth rate of beens was 
significantly (about 40\%) effected by 64 days exposure to 0.1 v/cm 
electric field at 45 Hz, when the been seeds were planted in soil. But no 
significant effect was observed when the soil was replaced with a nutrient 
solution.

Of course, possible genetic effects of VLF/ELF radiation are most interesting
in the context of our goals. Some early work suggested that weak electric and 
magnetic fields produced genetic aberrations in Drosophila, however these 
observations were not confirmed by subsequent experiments (Marino \& Becker 
\cite{marino}). Epidemiological evidence of possible carcinogenic effects of
electromagnetic field exposure is reviewed by Heath (\cite{heath}) and 
Davydov et al. (\cite{davydov}). It seems this field is subject to continuous
controversy. Some studies suggest that exposure to power frequency
electromagnetic fields may lead to increased risks of cancer, especially   
for leukemia and brain cancer. But other epidemiological studies did not reveal
any increased risk. For example, eight of the eleven studies conducted in
1991-1995 found statistically significant elevation of risk for leukemia. And 
four of the eight investigations that studied brain cancer also found some 
increase in risk (Heath \cite{heath}). Nevertheless  Heath considers the 
overall evidence as ``weak, inconsistent, and inconclusive''.

For energetic reasons, VLF/ELF radiation of not thermal intensity can not 
damage DNA or other cellular macromolecules directly. On this basis, the 
possibility that such weak electromagnetic fields can induce any biological 
effects was even denied for a long time (Binhi \& Savin \cite{savin}), until
a plethora of experimental evidence proved that ``Nature's imagination is 
richer than ours'' (Dyson \cite{dyson}). Let us mention one such recent
experiment of Tokalov et al. (\cite{tokalov}).

Cells have very effective emergency programs to cope adverse environmental 
conditions. Remarkably, cellular stress response is rather uniform 
irrespective to the stress factor nature. Some cellular functions that are 
not essential for survival, for example cell division, are temporarily 
suspended. Besides special kind of genes, the so called heat shock proteins
(HSP), will be activated. Their major function is the proper refolding of the 
damaged proteins. Heat shock proteins, notably the HSP70, were first 
discovered while investigating cellular responses to a heat shock, hence the 
name. Tokalov et al. (\cite{tokalov}) studied effects of three different 
stressors on the induction of several heat shock proteins and on the cell
division dynamics. The stress was produced by 200~keV X-ray irradiation, by 
exposure to a weak ELF electromagnetic field (50 Hz, $60\pm 0.2\; \mu T$), or
by a thermal shock ($41^\circ C$ for 30 min).

The pattern of induction of the most prominent members of the heat shock 
multigene family was found similar for all three stressors and HSP70 was the 
most strongly induced gene. But no effect on cell division was detected in
the case of ELF electromagnetic field exposure, in contrast with other two
stressors. Interestingly, when combined with the heat shock, ELF 
electromagnetic field shows cell protective effect: the number of 
proliferating cells strongly increases in comparison with the case when only
the heat shock stress is present. One might think that this protection
property is related to the induction of HSP70 genes by the electromagnetic 
field which helps to cope the thermal stress. But no protective effect was 
found when  ELF electromagnetic field exposure was combined with ionizing X-ray
irradiation. The reason of this difference is unknown, as are the molecular 
targets of ELF electromagnetic field. It was suggested that 
electromagnetic fields can act directly on DNA by influencing electron
transfer within the DNA double helix (Goodman \& Blank \cite{goodman}).  

The fact that weak electromagnetic fields can induce the stress proteins 
indicates that cells consider electromagnetic fields as potentially 
hazardous (Goodman \& Blank \cite{goodman}). This is surprising enough, 
because the magnitude of an effective magnetic stimulus is very small.
Electromagnetic fields can induce the synthesis of HSP70 at an energy 
densities fourteen orders of magnitude lower than heat shock (Goodman \& 
Blank \cite{goodman}). Such extra sensitivity to the magnetic field must
have good evolutionary grounds. Interesting thermo-protective effect of
the ELF electromagnetic field exposure mentioned above, and the absence of any
effects of weak electromagnetic fields on the cell proliferation, may indicate 
that cells are not really expecting any damage from the weak electromagnetic 
impulse, but instead they are using this impulse as some kind of early warning
system to prepare for the really hazardous other stress factors which often 
follow the electromagnetic impulse. There is another aspect of this problem
also: some recent findings in evolutionary biology suggest that heat shock 
proteins play important role in evolution.

HSP90 guides the folding process of signal transduction proteins which play
a key role in developmental pathways. When HSP90 functions normally, a large 
amount of genetic variation, usually present in genotype, is masked and does
not reveal itself in phenotype. However, under the stress Hsp90 is recruited 
to help chaperon a large number of other cellular proteins. Its normal role 
is impaired and it can no longer buffer variation. Therefore some mutations
will become unmasked and individuals with abnormal phenotype will appear in
the population. If a mutation proves to be beneficial in the new environmental
conditions, the related traits will be preserved even after the HSP90 resumes 
its normal function. Therefore HSP90 acts as a capacitor of evolution. If
environmental conditions are stable, the buffering role of HSP90 ensures the
stability of phenotype despite increased accumulation of hidden mutations in
genotype. When the environmental conditions suddenly change, as for example
after the asteroid impact, which is believed to cause the dinosaur extinction
65 million years ago, this great potential of genetic variation is released
in phenotype and the natural selection quickly finds the new forms of life 
with greater fitness. The Drosophila experiments of Rutherford and 
Lindquist (\cite{rutherford1}) demonstrated this beautiful mechanism, which
may constitute the molecular basis of evolution.

Further studies have shown that the HSP70 and HSP60 protein families also 
buffer phenotypic variation (Rutherford \cite{rutherford}). As was mentioned 
above, experiments demonstrated that ELF electromagnetic fields can induce
various heat shock proteins and in particular HSP70. Therefore we can 
speculate that ecological and genetic consequences of the Tunguska event are 
possibly not related to mutations which happened during the event, but are 
manifestations of the latent mutations, already present in the Tunguska biota,
which were unmasked due to the stress response. ELF/VLF radiation from the
Tunguska bolide might act as a stressor thereby explaining why the 
effect is concentrated towards the trajectory projection.
 
Note that direct mutagenic effect of the TSB flight and explosion is not also 
excluded. Because the Tunguska bolide was electrophonic bolide of the 
exceptional magnitude, very strong induced electric and magnetic fields are
expected, and therefore they could induce significant Joule heating in 
biological tissues. One can even find a witness accounts which can be 
interpreted as supporting this supposition, for example P.P. Kosolapov's 
report (Krinov \cite{krinov}):

``In June 1908, at about 8 in the morning I was in Vanavara settlement 
preparing myself for a hay harvest and I needed a nail. As I could not find it
in the hut, I went out in the yard and began to drag out the nail by pliers
from the window frame. Suddenly something as if burned my ears. Seizing them
and thinking that the roof was under fire, I raised my head and asked to 
S.B. Semenov who was sitting on the porch of his house: ``Did you see
something?'' - ``How not to see, answered he, I also felt as if I was embraced
with heat.'' After that I immediately went to the hut, but as I just entered it
and wanted to sit down on the floor to start work, a heavy blow followed,
soil began to drop from the ceiling, the door of the Russian stove was thrown  
out on the bed which stand in front of the stove and one window glass was 
broken. After that a sound like thunderclap appeared which receded in the north
direction. When there was  quiet again, I rushed out in the yard, but noticed
nothing any more.''

Krinov notes that the eyewitness did not mention any light phenomena and 
explains this by the fact that he was near the south wall of the hut and
thus shielded from the north half of the sky were the explosion took place.
Krinov further speculates that the heat sensation was caused by the bolide
glow as it fled overhead towards the explosion point. In our opinion more
realistic explanation is provided by the Joule heating due to extraordinary
strong electromagnetic pulse.

Many survived trees in the epicenter area have characteristic damages as if
originated from the lightning strikes. Therefore one can expect that the
explosion was accompanied by thousands of lightning strikes (Ol'khovatov 
\cite{olkhovatov}). It was proposed long ago that strong electric fields 
associated with thunderstorms could accelerate electrons to relativistic 
energies and lead to X-ray radiation. But all past attempts to register such 
radiation from lightnings have produced inconclusive results. At last, recent 
rocket-triggered lightning experiments unambiguously demonstrated that 
lightnings are accompanied by short intense bursts of ionizing radiation 
(Dwyer et al. \cite{dwyer}).

The detector used in these experiments (a NaI(Tl) scintillation counter) 
cannot distinguish between X-rays, gamma-rays and energetic electrons. So
the actual composition of the radiation burst is unknown, but the fact that
the radiation was not significantly attenuated by the 0.32 cm aluminum window
on the top of the detector ensures that the particle energies were much more
than 10 keV. The form of the observed signal indicates that the signal was
produced by multiple energetic particles. The bursts had typical durations 
of less than 100 microseconds and the total deposited energy was typically 
many tens of MeV per stroke. The energetic radiation seems to be associated 
with the dart leader phase of the lightning and precedes the return stroke
by about 160 $\mu s$.

Similar observations were made earlier by Moore et al. (\cite{moore}), who
observed energetic radiation from natural lightning. In this case the 
radiation burst was associated with much more slower stepped leader phase
and preceded the onset of the return stroke current by several milliseconds.

As we see, at present one has solid experimental evidence that lightnings are
sources of short bursts of ionizing radiation. Note that this experimental
fact cannot be explained by the conventional theories of high-voltage 
breakdown at high pressures and therefore they need to be revisited (Krider
\cite{krider}).

We do not know whether the TSB flight was also accompanied by ionizing 
radiation. This is not excluded as well because the strong electric fields 
associated  with the alleged space charge separation could produce energetic 
enough runaway electrons. Even if present, this radiation maybe will be too
attenuated before reaching the ground to produce significant biological 
effects. However, it seems very plausible that at least the explosion was 
accompanied by intense bursts of ionizing radiation from lightnings with
possible biological consequences. 
  
\section{The riddle of the sands}
We tried to argue in the previous sections that the genetic and ecological
impact of the Tunguska event is possibly related to the powerful ELF/VLF
electromagnetic radiation from the bolide and to the ionizing radiation
due to lightnings which accompany the explosion. Note that ionizing radiation
from the bolide and electromagnetic pulse as possible causes of genetic
mutations were considered earlier by Andreev and Vasilyev (Trayner 
\cite{trayner}) from different perspective. Turbulent wake behind the large 
enough bolide can produce 
required energetics of  ELF/VLF radiation, for example by Keay-Bronshten 
mechanism. The TSB was indeed very large, with the estimated prior to 
explosion mass between $10^5$ and $10^6$  tonnes (Trayner \cite{trayner}). 
The fact that no single milligram of this vast material was reliably 
identified in the epicenter region possibly tells against the asteroidal 
nature of the TSB (Bronshten \cite{bronshten1}). But cometary theory may also 
fail to explain the low altitude of the explosion, as well as some specific 
features of the forest devastation in the epicenter. These features indicate 
that besides the main explosion there were a number of lower altitude (maybe 
even right above the surface) less severe explosions (Vasilyev 
\cite{vasilyev}). The most striking fact is that the impression of the 
ballistic wave on the forest seems to extend beyond the epicenter of the 
explosion, as if some part of the Tunguska object survived the huge explosion 
and continued its flight (Vasilyev \cite{vasilyev}). Of course, it is a great 
enigma how an icy comet nucleus can lead to such strange effects. Maybe the 
key to this riddle is buried in the Libyan desert sands.

In 1932 an incredibly clear, gem-like green-yellow glass chunks were 
discovered in the  remote and inhospitable Libyan desert in western Egypt. 
Geologists dated the glass at 28.5 million years old and it is the purest 
natural silica glass ever found on Earth, with a silica content of 98\%. 
About 1400 tonnes of this strange material are scattered in a strewn field 
between sand dunes of the Great Sand Sea (Wright \cite{wright}).

The origin of Libyan Desert Glass (LDG) is not completely clear. LDG seems to
be chunks of layered tektite glass, the so called Muong Nong type of tektite
(M\"{u}ehle \cite{muehle}). Tektites are "probably the most frustrating 
stones ever found on Earth." (Faul \cite{faul}). The prevailing theory
about their origin is that they are formed from the rocks melted in large 
meteorite impacts. But secrets of glass making in such impacts are still 
unknown and some scientists even refuse that such high quality glasses, as 
one has, for example, in the LDG case, could ever originate as a result of a
fierce impact. The main argument against the terrestrial impact origin of 
tektites is the following (O'keefe \cite{keefe}). Tektites are unusually free
from the volatiles, like water and $\mathrm{CO_2}$, which are always present 
in terrestrial rocks. Glassmakers need several hours to remove bubbles
from the melted material to produce the high quality glass in industrial 
glass-making process. But the impact is very brief phenomenon, so there is not 
enough time to remove the volatiles.

O'keefe himself preferred lunar volcanism as an alternative explanation of 
tektite origin (Cameron \& Lowrey \cite{cameron}). In this approach tektites
are considered as ``Teardrops from the Moon'', in perfect agreement with
ancient legends (Kadorienne \cite{kadorienne}). Very romantic theory of 
course, but it encounters even more severe difficulties (Taylor \& Koeberl
\cite{taylor}). As a result the impact theory reigns at present.

But the glassmaker objection should be answered, and usually one refers to
shock compression (Melosh \cite{melosh}), the trick not used by glassmakers 
but expected in impact events. Due to shock compression at 100 Gpa, silicates 
almost instantaneously reach temperatures as high as 50000 $^\circ$C
(Melosh \cite{melosh}). Of course, nothing even remotely similar to such 
extreme conditions ever happens in industrial glass production. Therefore
the comparison is not justified.  

In the case of the Libyan Desert Glass, however, no impact crater have been 
found. Therefore Wasson and Moore (\cite{wasson}) suggested that an atmospheric
Tunguska-like explosion, $10^4$ times more powerful, was responsible for
LDG formation. Tremendous explosion heated 100km$\times$100km portion of the
entire atmosphere to temperatures high enough to melt small desert sand 
grains, which were elevated by generated turbulence. As a result a thin melt 
sheet of silicate was formed and a radiation background have kept it hot enough
for some time to flow and produce Muong Nong type tektites after 
solidification. Maybe multiple impacts produced by a fragmented comet, like
Shoemaker-Levy-9 crash with Jupiter, is needed to ensure the appropriate scale
of the event (Wasson \cite{wasson1}).

But the question about the high quality glass making reappears in this 
scenario, because now there are no extreme pressures associated with the 
impact cratering, and therefore no extreme compressive heating. Besides, 
evidence for shock metamorphism was revealed in some sandstones from the LDG 
strewn field by microscopic analysis (Kleinmann et al. \cite{kleinmann}).
This indicates to an impact, not to an atmospheric explosion. But then where 
is the crater? The situation is further involved by the recent strontium and 
neodymium isotopic study of these very sandstones and of some LDG samples 
(Schaaf \& M\"{u}ller-Sohnius \cite{schaaf}). Isotopic evidence indicates
the difference between the sandstones and LDG, therefore the former cannot
properly be regarded as possible source materials for LDG (Schaaf \& 
M\"{u}ller-Sohnius \cite{schaaf}).

As we see, Libyan Desert Glass, much like to the Tunguska event, suggests
very strange and peculiar impact. Maybe the required explanation should be
also very unusual and queer, like an impact of the mirror space body. Surely   
you will have a lot of glass after such an impact, won't you?

More seriously, the mirror matter idea is completely sound and attractive
scientific idea, which dates back to the Lee and Yang's (\cite{lee}) seminal
paper. This hypothetical form of matter is necessary to restore the 
symmetry between left and right. At the fundamental level the notions of left
and right (left-handed and right-handed spinors) originate because the Lorentz
group is locally identical to the SU(2)$\times$SU(2) group (see, for example,
Silagadze \cite{silagadze1}). Therefore one expects that the difference
between these two factors of the Lorentz group, the difference between left 
and right, should be completely conventional and the Nature to be left-right 
symmetric. But $\bf{P}$ and $\bf{CP}$ discrete symmetries are broken by the 
weak interactions, so they can not be used to represent the symmetry between
left and right, if we want a symmetric universe. One needs a new discrete 
symmetry $\bf{M}$, instead of charge conjugation $\bf{C}$, so that $\bf{MP}$
remains unbroken and interchanges left and right.

Lee and Yang (\cite{lee}) supposed that this new symmetry can be arranged if
for any ordinary particle the existence of the corresponding ``mirror" particle
is postulated. These new mirror particles are hard to detect because they
are sterile (or almost sterile) to the ordinary gauge interactions. Instead
they have their own set of mirror gauge particles, which we are blind to. The 
only guaranteed common interaction is the gravity (for a review and references
on mirror matter idea see Foot \cite{foot2}, \cite{foot3}, Silagadze 
\cite{silagadze2}, \cite{silagadze3}). Therefore big chunks of mirror matter
can be detected by their gravitational influence. This means that in the solar
system we do not have very much mirror matter, if any. But some amount is 
certainly allowed. Even a planetary or stellar mass distant companion to the 
sun is not excluded and represents a fascinating possibility (Foot \& 
Silagadze \cite{foot1}). 

Mirror matter is a natural dark matter candidate. It may even happen that the
mirror matter constitutes the dominant component of the dark matter
(Berezhiani et al. \cite{berezhiani}, Foot \& Volkas \cite{volkas}).
We know that there is a lot of dark matter in our galaxy and even in the solar 
neighborhood its density can reach roughly 15\% of the total mass density
(Olling \& Merrifield \cite{olling}). Therefore small asteroid size mirror 
objects occasionally colliding with the Earth is a possibility which can not 
be excluded.

What will happen during such collision depends on the details how the mirror
matter interacts with the ordinary matter. If the predominant interaction is
gravity, nothing interesting will happen, as the mirror asteroid will pass 
through the Earth unnoticed. But things will change if mirror and ordinary
matters interact via sizable photon-mirror photon mixing (Foot \cite{foot4},
Foot \& Yoon \cite{yoon}). In this case mirror charged particles esquire small
ordinary electric charge, they lose their sterility with respect to the
ordinary electromagnetic interactions and the mirror and ordinary nuclei will 
undergo Rutherford scattering, causing the drag force upon entry of a mirror
space body into the atmosphere.

In a number of detailed studies (Foot \cite{foot4}, Foot \& Yoon \cite{yoon},
Foot \& Mitra \cite{mitra}) the entry of a mirror body into the Earth's 
atmosphere was analyzed. The outcome depends on several factors, such as the 
magnitude of the photon-mirror photon mixing, the size of the mirror space 
body, its chemical composition and initial velocity. As concerned to the 
Tunguska problem, the most interesting conclusion was that a large ($R\sim 40~
\mathrm{m}$) chunk of mirror ice, impacting the Earth with initial velocity
of about 12 km/s, will not be slowed down much by the drag force in the 
atmosphere, but it will melt at a hight 5-10 km. Once it melts, the 
atmospheric drag force will increase dramatically, due to the body's expected 
dispersion, causing the body to release its kinetic energy into the atmosphere.
Therefore an atmospheric explosion is expected.

If the TSB was indeed a mirror asteroid or comet, as suggested by Foot 
(\cite{foot4}), the absence of the ordinary fragments is nicely explained.
Of course, mirror fragments are still expected, if the body had significant
non-volatile component. Maybe these fragments are still buried at the impact
site, but nobody bothers to dig them out.

Some other exotic meteoritic phenomena also appear less puzzling if one 
accepts that they were caused by mirror impactors (Foot \& Yoon \cite{yoon}).
And not only on the Earth. Looking at asteroid Eros and at its impact
craters, Foot and Mitra (\cite{mitra}) came to a strange conclusion that the 
small mirror space bodies in the solar system can actually outnumber the
ordinary ones. The reasoning is the following (Foot \& Mitra \cite{mitra}). 
When a mirror space body collides with an asteroid, it will release its 
kinetic energy at or below the asteroid's surface, depending on its size, 
velocity and the magnitude of the photon-mirror photon mixing. For Small 
mirror bodies the energy is released too slowly and over too large a volume 
to expect any crater formation. Therefore a crater hiatus is expected at some 
critical crater size, if the craters are caused by mirror impactors. And this 
is exactly what is observed for Eros: a sharp decrease in rate was found for
craters with diameter less than about 70 m.

Foot and Mitra (\cite{mitra}) was able to infer some estimation of the 
photon-mirror photon mixing magnitude from these observations. The result
fits nicely in the range which is expected if the anomalous meteoritic
phenomena, and the Tunguska event in particular, was indeed caused by the
mirror matter space bodies. There are some other interesting experimental
implications of the mirror matter which also involve the same range of the
mixing parameter (for a review, see Foot \cite{foot5}).  

Eros reveals still another footprint of the mirror world. Puzzling flat-flaw
crater ``ponds'' were unexpectedly discovered on its surface. The mirror
impact theory provides a ready explanation (Foot \& Mitra \cite{mitra}):
a large enough mirror space body while releasing its energy underground will 
melt surrounding rocks. If the photon-mirror photon parameter is negative, 
some extra heat is expected besides the kinetic energy. In this case mirror and
ordinary atoms attract each other, so for the mirror matter chunk it is 
energetically favorable to be completely embedded within ordinary matter,
releasing energy in the process (Foot \& Yoon \cite{yoon}).  

Interestingly, an enigmatic flat-flaw crater can be found even on Earth. It
is the nearly circular 38 km wide Rich\^at structure in Mauritania, the western
end of the Sahara desert. Nowadays this structure is no more considered as an 
impact structure, despite its uniqueness in the region, reported presence of 
coesite, and its round  "bull's-eye" shape. The reason is flat-lying strata at
the center of the structure with no signs of the disrupted and contorted beds,
lack of evidence for the shock-metamorphic effects, and the suspicion that the
reported coesite is in fact misidentified barite (Everett et al., 
\cite{everett}). The Rich\^at structure is believed to be a dome of endogenic 
origin sculpted by erosion. However why it is nearly circular remains 
a mystery.

Maybe Eros ponds hint the similar mirror matter explanation for this 
mysterious formation. Note that 50 km west-southwest of the Rich\^at structure
one finds similar but much smaller (of about 5 km diameter) formation -
the Semsiyat dome. If the ``soft'' impact of the main mirror body created
the Rich\^at structure, one can thought that its large fragment may caused the
Semsiyat dome.   

One expects numerous lightnings during a mirror impact event, because the
mirror space body will accumulate an ordinary electric charge while flying
in the atmosphere. The charge builds up because the ionized air molecules can 
be trapped within the mirror body, while the much more mobile electrons will 
escape (Foot \& Mitra \cite{mitra}). Interestingly, there were some 
speculations that the coesite-bearing quartzite breccias of the  Rich\^at  
structure  were  produced  by  lightning strikes (Master \& Karfunkel
\cite{master}). As already was mentioned above, there is an evidence that the
Tunguska event was accompanied by thousands of lightning strikes. More
recently, one can mention January 14, 1993 anomalous low altitude fireball
event in Poland, a candidate for a mirror meteorite strike, with enormous 
electrical discharge at the impact site, which destroyed most of the 
electrical appliances in nearby houses (Foot \& Yoon \cite{yoon}).

There is something similar to the Libyan Desert Glass in south Australia, the 
so called Edeowie glass field (Haines et al. \cite{haines}). Unlike to the
LDG, this enigmatic fused crustal material is quite clast-rich and 
inhomogeneous. The enigma in this case consists in the fact that no impact
crater has been found nearby, despite clear evidence that some rocks were
melted in situ. Haines et al. (\cite{haines}) concluded that lightning and
impact-related phenomena are the only reasonable possibilities capable
to produce the observed fusion. But in fact maybe there is one more reasonable
alternative, a large mirror body hitting the ground at cosmic velocity
(Foot \& Mitra \cite{mitra1}), capable to provide both the impact and
spatially and temporally confined intense lightnings. 

\section{Concluding remarks}
We have mentioned some riddles in this article, such as the Tunguska genetic
anomaly and electrophonic meteors, magnetoreception in biological systems and
molecular basis of evolution, Libyan desert Glass and Edeowie glass field,
flat-flaw Eros craters and the Rich\^at structure, parity violation and the
hypothetical mirror matter, and tried to argue that all these puzzles maybe
are just different pieces of the same one big jigsaw puzzle. The picture that
was assembled can be hardly considered as completely satisfactory, as in many
cases we rely on hypothesis instead of firmly established scientific facts.
Therefore we can not guarantee that the suggested picture is really the one
created by the jigsaw puzzle author (the Nature). Nevertheless it seems to
us interesting enough to offer to your attention.

As to the Tunguska genetic anomaly, we see the following picture as a 
reasonable explanation. The Tunguska bolide was of electrophonic nature. That 
means its flight was accompanied by a powerful ELF/VLF electromagnetic 
radiation. This radiation acted as a stress factor on the local biota and 
triggered subtle mechanisms to release the hidden genetic variations into the 
phenotype. Some direct mutagenic factors, related to the ionizing radiation 
associated with lightnings during the explosion, are also not excluded.

Interestingly, if the above given explanation is correct, the Tunguska genetic
anomaly represents in miniature the action of the molecular basis of evolution.
On much more grater scale, global catastrophic events, like the asteroid crash
65 million years ago which ended the dinosaur era, boost the evolution by the
same mechanism. We are left to admire the Grand Design of Nature and try to
survive its next evolutionary turn.

Therefore, finally let us return to outer space fears of sinful modern man.
For a long time the ancient belief that the cosmos can influence mundane
affairs was considered by scientists as a mere superstition. ``Modern
astronomers generally scoff at such superstitious beliefs, so it is somewhat
ironic that science has in the past few decades uncovered compelling evidence 
for celestial interference in terrestrial matters'' (Jewitt \cite{jewitt}). 
It is now widely accepted that near Earth objects larger than 1 km in diameter
represent considerable hazard and in the past Earth witnessed a number of
withering impacts, which maybe shaped the biological evolution. There are 
attempts to convince governments and society to fund ambitious projects
to completely identify potentially threatening near Earth objects and develop
adequate defense systems (Jewitt \cite{jewitt}).

But the Tunguska event and some other mysterious events of probably impact 
origin indicate the enigmatic type of soft impacts which do not leave any
crater, as well as impactor fragments, despite their tremendous magnitude.
In this article we mentioned one possible explanation that these impact events
are caused by mirror space bodies. Of course this explanation looks exotic,
but in fact it is the only one falsifiable in near future.

For mirror matter explanation of anomalous impact events the crucial ingredient
is the presence and magnitude of the photon-mirror photon mixing. And this
can be experimentally tested! In fact the crucial experiment is already 
planned. This is ETH-Moscow positronium experiment (Badertscher et al.
\cite{badertscher}, Foot \cite{foot5}). The photon-mirror photon mixing leads 
to the orthopositronium-mirror  orthopositronium oscillations. As a result in 
some tiny fraction of events the orthopositronium will decay into mirror 
photons and this will be detected as an event with missing energy. It is 
expected that the experiment will reach the needed sensitivity to prove or 
disprove the presence of the photon-mirror photon mixing of relevant 
magnitude (Badertscher et al. \cite{badertscher}, Foot \cite{foot5}).
   
If the ETH-Moscow positronium experiment outcome turns out to be positive, it
will mean bad news for fearful modern man, except the mirror matter theory 
proponents maybe. The peculiarities of the Eros craters, if really caused by
mirror impactors, indicate significant population of small mirror bodies in
the inner solar system. So the potential hazard for Earth is bigger than 
estimated. More importantly, it is very hard, if not impossible at all, 
to timely discover Earth approaching mirror space body and avert its impact. 
Therefore sinful modern man will be bound to face outer space hazards with 
eyes wide shut.

\begin{acknowledgements}
The author thanks V. Rubtsov, J. Moulder, A. Diamond, K. Thomson and S.B.Hagen
for sharing various information used in this article. He also appreciates
helpful comments from R. foot.
This research has made use of NASA's Astrophysics Data System, the high 
quality and usefulness of which is gratefully acknowledged.
\end{acknowledgements}

\end{document}